\documentstyle[12pt]{article}
\def\baselinestretch{1.5}
\begin{document}
\setcounter{page}{1}
\renewcommand{\thesection}{\Roman{section}}
\begin{center}
{\Large\bf Potential Harmonics Expansion Method for Trapped Interacting Bosons : Inclusion of Two-Body
Correlation
}\\

\vspace{1cm}
\normalsize\it

T. K. Das $^1$, B. Chakrabarti$^2$\\

$^1$
Department of Physics, University of Calcutta,
92 A. P. C. Road, Calcutta- 700009, India.\\
{\it{e-mail: tkdas@cubmb.ernet.in, tkdas6@hotmail.com}}\\
$^2$ Dept. of Physics and Astronomy, University of Oklahoma, Norman Ok
73019, U.S.A.\\ 
(present address: Department of Physics, K. N. College,
Berhampore 742101, W.B., India.)\\
{\it{e-mail: barnali@cubmb.ernet.in, barnali\_chakrabarti@hotmail.com}}\\
\end{center}
\rm
\begin{center}
{\bf ABSTRACT}
\end{center}  
We study a system of $A$ identical interacting bosons trapped by an
external field by solving {\it{ab initio}} the many-body Schr\"odinger
equation. A complete solution by using, for example, the traditional
hyperspherical harmonics (HH) basis develops serious practical problems
due to the large degeneracy of HH basis. Symmetrization of the wave
function, calculation of the matrix elements, etc., become an immensely
formidable task as $A$ increases. Instead of the HH basis, here we use a
new basis, called "potential harmonics" (PH) basis, which is a subset of
HH basis. We assume that the contribution to the orbital and grand orbital
[in $3(A-1)$-dimensional space of the reduced motion] quantum numbers
comes {\it{only from the interacting pair}}. This implies inclusion of
two-body correlations only and disregard of all higher-body correlations.
Such an assumption is ideally suited for the Bose-Einstein condensate
(BEC), which is required - for experimental realization of BEC - to be
extremely dilute. Hence three and higher-body collisions are almost
totally absent. Unlike the $(3A-4)$ hyperspherical variables in HH basis,
the PH basis involves only three {\it{active}} variables, corresponding to
three quantum numbers - the orbital $l$, azimuthal $m$, and the grand
orbital $2K+l$ quantum numbers for any arbitrary $A$. It drastically
reduces the number of coupled equations and calculation of the potential
matrix becomes tremendously simplified, as it involves integrals over
{\it{only three}} variables for any $A$. One can easily incorporate
realistic atom-atom interactions in a straight forward manner. We study
the ground and excited state properties of the condensate for both
attractive and repulsive interactions for various particle number. The
ground state properties are compared with those calculated from the
Gross-Pitaevskii (GP) equation. We notice that our many-body results
converge towards the mean field results as the particle number increases.\\[1cm]

PACS number(s): 03.65.Ge, 03.75.Hh, 03.75.Nt, 31.15.Ja\\


\newpage
\begin{tabbing}
{\Large{\bf I. Introduction}}
\end{tabbing}
\hspace*{3cm}
Although the phenomenon of Bose Einstein Condensation (BEC) was known
for a long time [1-3], its experimental observation in trapped and
supercooled (down to nano Kelvin temperatures)
alkali atoms in 1995 [4-6] renewed a great deal of interest - both experimental
and theoretical - in the phenomemon. The importance of this topic is clearly
demonstrated by the fact that two independent Nobel Prizes were awarded on
BEC related works in quick succession in the recent past. The density of
magneto-optically trapped atomic gas undergoing BEC is extremely low ( to
avoid recombination of atoms through three and higher body collissions) and
the number of trapped atoms
is typically of the order of a few hundred to a few million. This
is extremely small compared to the Avogadro number. For such a
small number of atoms an exact {\it{ab initio}} solution
would have been ideally desirable. But an interacting system of $A=(N+1)$
particles has $3N$ relative degrees of freedom and an {\it{ab initio}}
solution of the corresponding Schr\"odinger equation is practically impossible
for $A > 3$. Hence the usual theoretical tools that have been used so far
are the mean field models [7-10] and the Thomas-Fermi [8] approximation.
The dilute atomic gas undergoes BEC below a critical temperature ( typically
~ $10^{-9}$ degree K) when most of the atoms (bosons) go to the single particle
ground state. Then the de Broglie wavelength associated with the atomic
motion is much larger than the interaction length scale. Hence the resulting
many body system emerges as essentially a single quantum system where all the atoms
behave in a coherent manner [8,11]. At zero temperature, the effect of the excited states
are absent and the condensate is described by a single equation involving the
condensate wave function [8]. However this simple picture is no more true at a
finite temperature
due to the existence of interparticle interactions. The usual procedure is to
start with the mean field approximation like the Hartree-Fock (HF) theory
for the many body system [7-10]. This is an independent particle approach where each individual atom
is assumed to move in a single particle orbit. These orbits are
determined self consistently by allowing an atom in one orbital to be influenced
by other atoms in other orbitals through two-body interaction. Assuming a
contact interaction for the two-body potential, {\it{viz.}},
$V(\vec{r} - \vec{r^{\prime}})$ = $g \delta(\vec{r} - \vec{r^{\prime}})$,
the many body equation reduces to the famous Gross-Pitaevskii (GP)
equation [8]. At zero temperature, the effect of excited states are neglected
and the condensate is described by the time independent GP equation
\begin{equation}
\left[ - \frac{\hbar^{2}}{2m} \bigtriangledown ^{2} + V_{ext}(\vec{r})
+ g\phi^{2}(\vec{r})
\right] \phi(\vec{r}) = \mu \phi(\vec{r})
\end{equation}
where $n(\vec{r})$ = $\phi^{2}(\vec{r})$ is the condensate density and $\mu$
is the chemical potential.
For a first approach the contact interaction is justified since in the
cold and dilute gas only binary collissions at low energies are relevant.
These are characterized by the $s$-wave scattering length $(a_{sc})$, which
is independent of the details of  two-body potentials. The strength constant
$g$ of the contact interaction is
related to the scattering length through [8] 
\begin{equation}
g= \frac{4\pi\hbar^{2}a_{sc}}{m}
\end{equation}
The GP equation has been used extensively to study the BEC [8,11]. Although
most of the static, dynamic and thermodynamic properties are fairly
well reproduced by the GP equation [8], the wave function does not include
any correlation. Furthermore the assumption of a contact $\delta$-interaction
is too simple and does not represent the realistic situation. It has already
been shown that the Dirac $\delta$-function is not suitable as a replacement of the
actual two-body interaction in exact theories in more than one dimension 
[12]. This is because the
Hamiltonian then becomes unbound from below and the ground state energy
diverges for an attractive zero range potential. Solutions are usually
obtained in the metastable region (although such solutions are not
rigorously correct for an attractive $\delta$-function potential) and the
condensate becomes unstable for $N$ larger than a critical number, due to
disappearance of the local minimum. This was shown by Bohn {\it{et al}} in a 
hyperspherical calculation keeping the lowest (most dominant) harmonic [13].
A third disadvantage is the
non-lineraity of the GP equation, so that standard quantum mechanics is
not applicable without concessional approximation.
Thus one has to go beyond the mean field approximation and simple contact interactions.\\
\hspace*{3cm}
Because of the limitations of the mean field theory and GP
equation it is desirable to solve the many body {\it{linear}} Schr\"odinger 
equation directly.
The Schr\"odinger equation for a
system of $A=(N+1)$ identical bosons, each of mass $m$, confined by
an external field $V_{trap}^{\prime}$ (acting on each individual boson) and interacting
through a mutual two body interaction $V$ is
\begin{equation}
\left[- \frac{\hbar^{2}}{2m} \sum_{i=1}^{A}\bigtriangledown_{i}^{2}
+ \sum_{i=1}^{A} V_{trap}^{\prime}(\vec{x_{i}})
+\sum_{i<j=2}^{A}V(\vec{x_{i}}-\vec{x_{j}})\right]
\Psi(\vec{x}) = E \Psi(\vec{x})
\end{equation}
where $\vec{x}$ refers to the set of particle coordinates
$\{\vec{x_{1}}, \vec{x_{2}}, ......\vec{x_{A}} \}$ of $A$ bosons.
The center of mass (CM) motion can be eleminated resulting in a Schr\"odinger equation
in $3N$ variables. A standard practice is the use of hyperspherical harmonics
expansion (HHE) method,  in which the wave function is expanded in the complete
set of hyperspherical harmonics (HH) spanning the $(3N-1)$-dimensional
hyperangular space [14]. Projection on a particular HH leads to a system of coupled
differential equations (CDE). However there are several very serious difficulties
associated with the solution of a fairly large number of particles.
Firstly the expansion basis of HH should be properly symmetrized and appropriate
conserved quantum numbers properly taken care of.
Secondly calculation of matrix elements of all the pairwise
two-body potentials is an extremely formidable task. Finally,
due to very large degeneracy of the HH basis for a large number of particles,
the number of CDE and the dimemsion of the potential matrix is too large to
be handled by any computer [14].
On top of all these, the convergence rate of the HH expansion, especially
for long-range interactions, is slow [15].
For these reasons the HHE method has been used fully for the three body system
only [15-18]. 
On the other hand, as we discussed earlier, the condensate can be treated broadly 
as a "single lump of quantum stuff",  since all the individual atoms in the condensate lie
within one single de Broglie wavelength [8]. Thus it is reasonable to assume
that the basic properties of the condensate in the lowest approximation, is described
by a single collective coordinate. This led Bohn {\it{et. al.}} [13] to go
for the K-harmonic approximation, in which the HH expansion is restricted
effectively to the first term only ( which is independent of the hyperangles).
Such a {\it{drastic}} approximation may be justified for a contact interaction
only. Even in this case, for an attractive $\delta$-function interaction,
there are no {\it{rigorously stable}} solutions. Since the wave function becomes
independent of the hyperangles and the hyperradius is invariant under any permutation of
the particles, the wave function becomes totally symmetric, as required.
The calculation of the potential matrix also simplifies immensely and the
CDE reduces to a single differential equation [13].
The hyperradius emerges as the sought for collective coordinate. In spite of the
great simplifications, there are serious criticisms of this approach :
(1) The method cannot be applied to any realistic two-body interaction.
(2) Even for a contact interaction, the method is not satisfactory for attractive
$\delta$-function interaction, for which no rigorous solution extsits. 
(3) Only one collective variable is involved. Hence
it can only describe the gross features of the condensate, without any finer
details. Thus a more rigorous treatment is necessary. But as already mentioned
a completely rigorous, essentially exact solution of the Schr\"odinger
equation is possible for the three body system only. That has been done
to get an idea of the initial trend as the particle number increases
from three by Esry and Greene [12]. However that is far from the real 
situation in a condensate. \\
\hspace*{3cm}
An alternative approach of exact numerical diagonalization of the many
body Hamiltonian was adopted by Haugset and Haugerud [19] for a small
number ($\leq$ 30) of interacting (via contact interaction) bosons confined by a harmonic trap.
However, this was restricted to {\it{one and two dimensions only}}.
Moreover the process is extremely time consuming {\it{even for two
dimensional}} condensates, with a nagging question of convergence of the
chosen harmonic oscillator basis expansion. The rate of convergence is
expected to be slower for a realistic two-body interaction and in three
dimensional condensates. Although analytic expressions for the matrix elements 
are greatly simplified for a delta function interaction, all the problems 
associated with a contact interaction discussed above remain for the two 
dimensional condensate. However, there is no problem with the one dimensional 
condensate, as one dimensional delta function is not pathological.\\
\hspace*{3cm}
From the above discussion it is clear that an exact treatment of the many
body system in three dimensions 
is not possible beyond the three body system. On the other hand the
single quantum nature of the entire condensate suggests that out of the
thousands to millions of degrees of freedom of the individual particles
{\it{ only a few}} are physically relevant. This is due to the fact that the
condensate is possible only at extremely low temperatures ( low energy of the
individual particles) and extremely low densities. Under these conditions
{\it{ only two body collisions}} are relevant. Three and higher body
collsions are {\it{extremely}} rare and correlations beyond two body
correlations in the condensate wave function are completely negligible upto
a very high degree of precision. Indeed in an experimental situation this
is ensured by keeping the density extremely low, so that there are no
recombination via three and higher body collisions [8]. The mean field
approach ignores all correlations including two-body correlations.
Importance of two-body
correlations in BEC has been emphasized by several authors [20,21]. Thus 
physically relevant quantities
are contributed by two-body collisions, while the rest of the particles
in the condensate do not partate in any motion other than a collective one and
are simply inert spectators. The emerging picture then suggests that 
most of the degrees of freedom
of these spectators can be frozen, while a single pair interacts.
This reduces the physically important degrees of freedom of the condensate
to just four - a global length scale (hyperradius) of the entire condensate,
and the three degrees of freedom of the relative vector $\vec{r_{ij}}$ =
$\vec{x_{i}}-\vec{x_{j}}$ of the interacting pair. However one has to concede 
that {\it{any}} pair out of the $A=(N+1)$ atoms in the condensate can interact.
These are also consistent with the intuitive "single quantum stuff" concept
of the condensate. \\
\hspace*{3cm}
Among the various possible theoretical approaches to handle the many body
system, the HHE method appears to be the most lucrative one, as it readily
provides the hyperradius as the most important collective variable. A 
theoretical formalism, arising out of the HHE method, was adopted by
Fabre de la Ripelle [22] in 1986. Although the primary concern there was
an application to the nuclear systems consisting of fermions, it was noted
that the formalism
is applicable to a system of identical bosons also [23]. To incorporate 
the importance of the interacting pair and two-body correlations,
he introduced the potential harmonics (PH) expansion basis [23],
rather than the general HH basis, thereby reducing the expansion basis to a
great extent. Potential harmonics is a subset of HH, where all
correlations higher than two-body ones are disregarded. In PH, the
contribution to the total orbital angular momentum as also the grand
orbital quantum number comes {\it{only}} from the interacting pair. Here
all the ($A-2$) spectators are assumed to be described by the HH of the
lowest (zero) order. We adopt this procedure since this approximation 
is quite justified in our
situation due to the diluteness of BEC, where two-body correlation is the
most important and all higher-body correlations can be safely ignored.
Using Faddeev like decomposition of the total wave function, and then
expanding each such component in an appropriate set of PH, the number of
CDE can be reduced drastically. Since the PH involves only four active
degrees of freedom, calculation of potential matrix elements is simplified
tremendously as compared to that in HH basis. Use of realistic 
two-body interactions and calculation of their matrix elements are quite
straight forward. Requiring the Faddeev component for the
($ij$) interacting pair to be symmetric under ($ij$)-pair exchange, the
total wave function becomes automatically totally symmetric. Thus the
symmetrization of the wave function is also handled properly.\\
\hspace*{3cm}
Thus a truely many body equation is reduced to a tractable
mathematical form. The assumptions leading to this are especially appropriate
for the BEC. Hence we adopt the PH basis as our starting point. This is 
theoretically applicable to a system containing any number of particles, 
but we will see in Sec. III, 
that numerical difficulties
arise as the number of particles increases beyond a certain number. In this
communication we report some of the basic properties of the condensate
for various particle numbers and compare them with previous calculations.\\
\hspace*{3cm} Sorensen {\it{et al}} [20,21] have followed a method which
is similar in spirit to the present work, although it differs in details. 
They expand the wave function in the adiabatic subset
{$\Phi_{n}(\rho,\Omega)$} of the {\it{full}} $(N-1)$-body Hamiltonian
(in CM frame). Later this is decomposed in Faddeev like
components $\phi_{ij}$. This leads to an integro-differential equation
(IDE) for $\phi$ (=$\phi_{ij}$, which is the same for {\it{all}}
$ij$-pairs due to boson symmetry) involving {\it{five}} dimensional
integrals and the full $(3N-4)$-dimensional hyperangular
differential operator $\hat{\Lambda^{2}}$. 
All $(3N-5)$ angle derivatives other than $\alpha = \alpha_{12}$
(where $r_{ij} = \sqrt{2} \rho sin \alpha_{ij}$ is the relative separation
of the ($ij$)-pair and $\rho$ is the hyperradius of the full system), are 
disregarded, leaving only one angle variable. Assumption of
a {\it {very short ranged}} two-body potential reduces the five
dimensional integrals to two dimensional ones. In this limit simple
expressions are obtained for the integrals in IDE.
On the other hand, we
write the complete $3N$-dimensional Sch\"odinger equation of the relative
motion of a $(N+1)$ boson system in terms of Faddeev like 
components $\Phi(\vec{r_{ij}},r)$, 
subject to the approximation that $\Phi(\vec{r_{ij}},r)$
corresponds to zero eigenvalue of the hyper angular momentum operator (see
later) for
the $(N-1)$ remaining relative vectors of the spectators, while
$(ij)$-pair interacts. These are then expanded in the potential
harmonics (PH) basis. The assumptions in our method are
clearly justified in terms of the physics of the chosen system, which 
have been stated earlier. 
While the use of PH basis in nuclei (as originally used by
Fabre in [22,23]) is questionable due to high spatial density of
nucleons in a nucleus, its application in BEC is ideally suited (the
number of atoms in the condensate is $\leq 10^{6}$ in a space of
macroscopic linear dimensions of order $10^{-2} cm$, which  is immensely
smaller than the Avogadro number). As a consequence, the total orbital
($l$) and grand orbital ($K$) angular momenta of the system are
contributed by the interacting pair alone. Apart from this well justified
fundamental approximation, we need no other approximation.
Although for the first calculation, we have restricted ourselves to $l$ =
0 and a {\it {central}} two-body interaction, both these can be relaxed
resulting in a somewhat more complicated equation. Finally
the system of coupled differential equations in one variable
(hyperradius, $r$) can be solved numerically, {\it {without additional
approximation}} (as done in ref. [15] and compared with adiabatic
approximation (AA) in ref. [28]) using, {\it{e.g.}} renormalized Numerov
method. Once again, as a preliminary calculation, we
use AA to solve the CDE. Our use of AA in solving the CDE is not an 
indispensable one; it is done only to reduce the numerical 
calculation. But in the approach of Sorensen {\it{et al}}, adiabatic 
subset is the starting point to separate the hyperangular and 
hyperradial motions. Furthermore our method can handle any two-body 
potential (central or not, short ranged or not); for non-central 
potential, calculation of matrix elements will involve integrals over 
two polar angles in addition. The approach of ref. [20,21] requries 
a very short ranged, central potential to reduce the equation to a 
manageable form. The present method has no such restriction.\\
\hspace*{3cm}
The paper is organised as follows. In Sec. II, we present our choice of
Jacobi coordinates and express the kinetic energy in the chosen set of
hyperspherical variables. In the same section, we introduce the concept of
potential harmonics basis and obtain the set of coupled differential
equations resulting from the many-body Schr\"odinger equation. The
numerical method for solving the CDE and results of our calculation are
presented in Sec. III. There we compare our results  
for different numbers of particles with those of
earlier calculations. Finally in Sec. IV we draw our conclusions. Some of
the detailed expressions have been given in the Appendix.\\
 
\begin{tabbing}
{\Large{\bf II. Theory}}
\end{tabbing}
\begin{tabbing}
{\large{\bf A. Choice of Jacobi coordinates}}
\end{tabbing}
\hspace*{3cm}
We consider a system of $A$ = $(N+1)$ identical bosons, each of mass $m$ and 
confined magnetically in a trap which is approximated by a spherically symmetric 
harmonic oscillator potential with frequency $\omega$. We assume that the atomic cloud is 
at zero temperature. The full many body Hamiltonian is given by 
\begin{equation}
\left[ -\frac{\hbar^{2}}{2m} \sum_{i=1}^{N+1} \bigtriangledown _{i}^{2} + 
\sum_{i=1}^{N+1}\frac{1}{2}m\omega^{2}x_{i}^{2} + \sum_{ij>i}^{N+1}V(\vec{x_{i}}-\vec{x_{j}})\right]
\Psi(\vec{x})=E^{\prime} \Psi(\vec{x})
\end{equation}
where $\vec{x}$ refers to the set of particle coordinates
$\{\vec{x}_{1}, \vec{x}_{2},
 ..., \vec{x}_{N+1}\}$ of $(N+1)$ bosons and $E^{\prime}$ is the total energy.
We decompose the total wave function $\Psi(\vec{x})$ as the sum of pairwise 
partial waves 
\begin{equation}
\Psi(\vec{x}) = \sum_{ij>i}^{N+1} \psi_{ij}(\vec{x})
\end{equation}
The Schr\"odinger equation for $\psi_{ij}$ can be written as 
\begin{equation}
(T+V_{trap}^{\prime}-E^{\prime})\psi_{ij}(\vec{x})=-V(r_{ij})\sum_{kl>k}^{N+1} \psi_{kl}(\vec{x})
\end{equation}
where $T$ is the total kinetic energy operator, $V_{trap}^{\prime}$ is the
confining potential,  
$V_{trap}^{\prime}$ = $\sum_{i=1}^{N+1}\frac{1}{2} m \omega^{2} x_{i}^{2}$ and $V(r_{ij})$
is the pairwise local central two-body interaction between $i^{th}$ and
$j^{th}$ particles, $\vec{r_{ij}}= \vec{x_{i}}-\vec{x_{j}}$. Applying the operator $\sum_{ij>i}^{N+1}$ on both sides 
of eq.(6), and using eq.(5), we get back eq.(4). Now instead of $(N+1)$
particle coordinates $\vec{x_{i}}$, the system can alternatively be described by 
the center of mass coordinate $\vec{R}$  
\begin{equation}
\vec{R} = \frac{1}{N+1} \sum_{i=1}^{N+1}\vec{x_{i}}
\end{equation}
and $N$ Jacobi coordinates defined as 
\begin{equation}
\vec{\zeta_{i}} = \sqrt{\frac{2i}{i+1}} \left(\vec{x_{i+1}}-\frac{1}{i}\sum_{j=1}^{i}
\vec{x_{j}} \right), \hspace*{.5cm} i= 1, ..., N \hspace*{0.3cm}.
\end{equation}
The chosen normalization of $\vec{\zeta_{i}}$ facilitates writing the Laplace operator
in the form
\begin{equation}
\frac{1}{2} \sum_{i=1}^{N+1} \bigtriangledown _{i}^{2} = \frac{1}{2A} \bigtriangledown_{R}^{2}
+ \sum_{i=1}^{N} \bigtriangledown_{\zeta_{i}}^{2} \hspace*{0.3cm}.
\end{equation}
Then the relative motion (after removal of center of mass motion from
eq.(4)) is described  
by [14,23]
\begin{equation}
\left[-\frac{\hbar^{2}}{m} \sum_{i=1}^{N} \bigtriangledown_{\zeta_{i}}^{2}+
V_{trap}
+ V_{int}(\vec{\zeta_{1}}, ..., \vec{\zeta_{N}})-E \right] \psi(\vec{\zeta_{1}},
 ..., \vec{\zeta_{N}}) = 0 \hspace*{0.3cm},
\end{equation}
where $V_{trap}=\sum_{i=1}^{N}\frac{1}{2}m\omega^{2}\zeta_{i}^{2}$ and
$V_{int}$ is the sum of all pairwise interactions, 
$V_{int}=\sum_{ij>i}^{N+1}V(r_{ij})$ expressed in the relative coordinates.
Here $E$ is the energy of the relative motion, {\it{i.e.}}, $E^{\prime}$
minus energy of CM motion.
The hyperradius $r$ is defined as [22]
\begin{equation}
r= \left[\sum_{i=1}^{N} \zeta_{i}^{2} \right]^{1/2} = \left[\frac{2}{N+1} 
\sum_{i,j>i} r_{ij}^{2} \right]^{1/2} \hspace*{0.3cm},
\end{equation}
which is invariant under permutations of the particle indices as also
three dimensional rotations. The hyperspherical
coordinates are constituted by the hyperradius $r$ and remaining $(3N-1)$ 
hyperangles, denoted collectivelty by $\Omega_{N}$ in $D$ = $3N$
dimensional space.  Note that the choice of Jacobi coordinates eq.(8), is not unique, since the labelling of the particle indices and consequently that of the 
Jacobi coordinates are arbitrary. We choose a particular set by specifying
the relative separation of the interacting pair, $\vec{r_{ij}}$
as $\vec{\zeta_{N}}$ and $(\vartheta, \varphi)$ are the two spherical
polar coordinates associated with $\vec{r_{ij}}$.
The relative length is defined in terms of $\phi$ 
through $r_{ij} = r \hspace*{.1cm}cos\phi$.
For the rest of $(N-1)$ Jacobi coordinates, we define
the hyperradius $\rho_{ij}$ in the $3(N-1)$ dimensional space by
\begin{equation}
\rho_{ij} = [\sum_{k=1}^{N-1}\zeta_{k}^{2}]^{1/2}
\end{equation}
which is related with $\vec{\zeta_{N}}$ = $\vec{r_{ij}}$ by 
\begin{equation}
\rho_{ij}^{2} + r_{ij}^{2} = r^{2}, \hspace*{1cm} \rho_{ij} = r \hspace*{.1cm}
sin\phi \hspace*{1cm} \cdot
\end{equation}
Then our hyperspherical coordinates become
\begin{equation}
(r,\Omega_{N}) = (r,\phi,\vartheta, \varphi, \Omega_{N-1}) \hspace*{1cm} \cdot 
\end{equation}
Here $\Omega_{N-1}$ involves $2(N-1)$ spherical polar angles
associated with each of $(N-1)$
Jacobi vectors $\{\vec{\zeta}_{1},\vec{\zeta}_{2}, ...,
\vec{\zeta}_{N-1}\}$ 
and $(N-2)$ angles (expressing relative lengths) , i.e., a
total of $(3N-4)$ variables. In this choice of
hyperspherical coordinates, the Laplace operator takes the form [22]
\begin{equation}
\bigtriangledown^{2} \equiv \sum_{i=1}^{N} \bigtriangledown_{\zeta_{i}}^{2}=
\frac{\partial^{2}}{\partial r^{2}} + \frac{3A-4}{r}
\frac{\partial}{\partial r} + \frac{L^{2}(\Omega_{N})}{r^{2}}, \hspace*{.5cm} A= N+1
\end{equation}
$L^{2}(\Omega_{N})$ is the grand orbital operator in $3N$ dimensional space
which is obtained
from a recurrence formula [22] and has the form
\begin{equation}
L^{2}(\Omega_{N}) = 4(1-z^{2}) \frac{\partial^{2}}{\partial z^{2}} +
6 [ 2-N(1+z)] \frac{\partial}{\partial z} + 2 \frac{l^{2}(\omega_{ij})}{1+z} 
+ 2 \frac{L^{2}(\Omega_{N-1})}{1-z} 
\end{equation}
where $z= cos2\phi$, $\omega_{ij}$ reprsents the two polar angles
$(\vartheta,\varphi)$
associated with $\vec{r_{ij}}$ and  $L^{2}(\Omega_{N-1})$ is the
grand orbital operator in
$3(N-1)$ dimensional space. 
\begin{tabbing}
{\large{\bf B. Potential basis and potential multipoles}}
\end{tabbing}
\hspace*{3cm}
To exapand a function $V(r_{ij})$ in hyperspherical harmonics (HH) we use the above definition
of Jacobi coordinates. It is easy to see that HH basis which is complete for the expansion
of $V(r_{ij})$ does not contain any function of the coordinate
$\vec{\zeta_{i}}$ with $i<N$ and
is given by [23]
\begin{equation}
{\mathcal P}_{2K+l}^{l,m} (\Omega_{ij}) = Y_{l}^{m}(\omega_{ij})\hspace*{.2cm}
^{(N)}P_{2K+l}^{l,0}(\phi) {\mathcal Y}_{0}(D-3)
\end{equation}
where $^{(N)}P_{2K+l}^{l,0}$ is a function involving the Jacobi polynomial
and is needed in the general expression of the hyperspherical harmonics
(see Appendix) of grand orbital $2K+l$ and orbital angular momemtum $l$. The
quantity
${\mathcal Y}_{0}(D-3)$ is the HH of order zero (i.e. grand orbital quantum number is zero)
in $3(N-1)$ dimensional space,
${\mathcal Y}_{0}(D-3)$ = $\left( \frac{\Gamma ((D-3)/2)}{2\pi^{(D-3)/2}}\right)^{1/2}$.
This new basis set which is a subset constituted by HH of order $(2K+l)$ are 
called "potential harmonics" (PH). These are the eigenfunctions
of $L^{2}(\Omega_{N})$,
when the eigenvalue of $L^{2}(\Omega_{N-1})$ is 0 and satisfy the
eigenvalue equation :
\begin{equation}
\left[L^{2}(\Omega_{N}) + {\cal L}({\cal L}+D-2) \right]
{\mathcal P}_{2K+l}^{l,m}(\Omega_{ij}) =0, \hspace*{1cm} {\cal L} = 2K+l
\hspace*{.5cm} \cdot
\end{equation}
The relation $L^{2}(\Omega_{N-1})\psi_{ij}(\vec{x}) = 0$ implies
that we are considering only those states which are
invariant under all generalized rotations in $3(N-1)$ dimensional space. Natuarally
the contribution
to the grand orbital quantum number comes only from the interacting pair.
This corresponds effectively to
two-body correlations only in the wave function.
Due to diluteness of atomic BEC, the effect of higher body correlations
can be ignored as the probability for three or more particles to come
close at the
same time is extremely small. This reduces the number of quantum numbers
in the new basis (all the quantum numbers specifying the eigenfunctions
of $L^{2}(\Omega_{N-1})$ are zero). It
contains only three quantum numbers; orbital $l$, azimutal $m$ and grand orbital $2K+l$ for any $N$, 
instead of $(3N-1)$ quantum numbers corresponding to $(3N-1)$
hyperspherical variables in the general HH basis.
The normalization condition is given by
\begin{equation}
\int { {\mathcal P}_{2K+l}^{{l,m}^{*}}(\Omega_{ij})
{\mathcal P}_{2K^{\prime}+l^{\prime}}^{l^{\prime} m^{\prime}}(\Omega_{ij})} d\Omega_{ij}
=\delta_{KK^{\prime}} \delta_{ll^{\prime}}\delta_{mm^{\prime}} \hspace*{1cm} \cdot 
\end{equation}
Then the PH expansion of the potential is
\begin{equation}
V(r_{ij}) = \sum_{K,l,m} A_{l}^{m}(i,j) {\mathcal P}_{2K+l}^{l,m}(\Omega_{ij})V_{K}^{(D,l)}(r)
\end{equation}
$A_{l}^{m}(i,j)$ is an operator which is independent of $r_{ij}$, 
but may act on other variables like spin variables.
The quantity $V_{K}^{(D,l)}(r)$ are the "potential multipoles" and for a
central potential, it is given by [23] 
\begin{equation}
\begin{array}{rcl}
V_{K}^{(D,l)}(r) &=& <{\mathcal P}_{2K+l}^{l,m}(\Omega_{ij})|V(r_{ij})>\\
&=& |{\mathcal Y}_{0}(D-3)|^{-1} \int_{0}^{\pi/2} {^{(N)}P_{2K+l}^{l,0}(\phi)}
V_{l}(rcos\phi)(sin\phi)^{D-4}(cos\phi)^{2}d\phi \hspace*{0.3cm},\\
\end{array}
\end{equation}
where the functions $^{(N)}P_{2K+l}^{l,0}(\phi)$ are defined in the Appendix.
Starting from the multipoles calculated either for the $D=5$ or $D=6$ ( depending wheather $D$ is odd 
or even) and using simple recurrence formul{\ae} 
potential multipoles for any
D can be calculated [23]. 
\begin{tabbing}
{\large{\bf C. Coupled differential equations}}
\end{tabbing}
Splitting eq.(10) in the manner of eq.(6) for the $(ij)$-interacting pair
and using eqs.(14)-(16), subject to the restriction that the eigenvalue of
$L^{2}(\Omega_{N-1})$ is zero, we see that the $(ij)$ Faddeev component
will be a function of $\vec{r}_{ij}$ and $r$ only and satisfies [23] 
\begin{equation}
(T+V_{trap}-E) \Phi(\vec{r_{ij}}, r)
= - V(r_{ij}) \sum_{k,l>k} \Phi(\vec{r_{kl}},r) \hspace*{0.3cm},
\end{equation}
where $\Phi(\vec{r_{ij}},r)$ differs from the general solution $\psi_{ij}$
by the fact that it corresponds to eigenvalue zero of the operator
$L^{2}(\Omega_{N-1})$.
Next expand the wave function $\Phi(\vec{r_{ij}},r)$ in the
complete set of potential
harmonics (when $l$ is a good quantum number) as 
\begin{equation}
\Phi(\vec{r_{ij}},r) = r^{-\frac{D-1}{2}} \sum _{K^{\prime}} 
{\mathcal P}_{2K^{\prime}+l}^{lm}(\Omega_{ij}) u_{K^{\prime}}^{l}(r)
\end{equation}
Substitution of eq.(23) in eq.(22) and projection on the same basis,
leads to the set of coupled differential equations [23]
\begin{equation}
\left[-\frac{\hbar^{2}}{m}\frac{d^{2}}{dr^{2}}+\frac{\hbar^{2}}{m}
\frac{{\cal L}_{K}({\cal L}_{K}+1)}{r^{2}} + V_{trap}(r) -E \right] u_{K}^{l}(r) +
\sum_{K^{\prime}} f_{K^{\prime}l}^{2}V_{KK^\prime}(r)u_{K^{\prime}}^{l}(r)=0
\end{equation}
where 
\begin{equation}
\begin{array}{rcl}
{\cal L}_{K} &=& 2K+l+\frac{D-3}{2}\\
f_{Kl}^{2} &=& \sum_{k,l>k} <{\mathcal P}_{2K+l}^{lm}(\Omega_{ij})|{\mathcal P}_{2K+l}^{lm}
(\Omega_{kl})>\\
\end{array}
\end{equation}
The potential matrix is given by
\begin{equation}
V_{KK^{\prime}}(r) = \int {\mathcal P}_{2K+l}^{{lm}^{*}}(\Omega_{ij})V(r_{ij})
{\mathcal P}_{2K^{\prime}+l}^{lm}(\Omega_{ij})d\Omega_{N}
\end{equation}
So instead of $(3N-1)$ angle variables in HHE method, in potential harmonics
expansion method (PHEM) the integral invloves
only 3 angle variables.
It greatly simplifies the calculation of the matrix element for any $N$.\\
\hspace*{3cm}
The quantity $f_{kl}^{2}$ of eqs. (24) and (25) is given by [23]
\begin{equation}
f_{Kl}^{2}= 1+ [2(A-2)(-\frac{1}{2})^{l}P_{K}^{\alpha\beta}(-\frac{1}{2})
+\frac{(A-2)(A-3)}{2}P_{K}^{\alpha\beta}(-1)\delta_{l,0}]/P_{K}^{\alpha\beta}(1)
\end{equation}
where $\alpha$ = $(3A-8)/2$ and $\beta$ = $l+\frac{1}{2}$ and
$P_{K}^{\alpha\beta}(x)$ is the Jacobi polynomial.
Multiplying eq. (24) by appropriate constant factors, it can be put in a
symmetric form:
\begin{equation}
\begin{array}{cl}
&\left[-\frac{\hbar^{2}}{m} \frac{d^{2}}{dr^{2}} + \frac{\hbar^{2}}{mr^{2}}
\{ {\cal L}({\cal L}+1) + 4K(K+\alpha+\beta+1)\}+V_{trap}(r)-E\right]
U_{Kl}(r) \\
+ & \sum_{K^{\prime}}\overline{V}_{KK^{\prime}}(r) U_{K^{\prime}l}(r) = 0
\end{array}
\end{equation}
where ${\cal L}$ = $l$ + $(3A-6)/2$, the symmetrized
potential matrix $\overline{V}_{KK^{\prime}}$ has the form
\begin{equation}
\overline{V}_{KK^{\prime}}(r) = f_{Kl}V_{KK^{\prime}}(r)f_{K^{\prime}l}(h_{K}^{\alpha\beta}
h_{K^{\prime}}^{\alpha\beta})^{-\frac{1}{2}}
\end{equation}
and
\begin{equation}
U_{Kl}(r) = f_{Kl}(h_{K}^{\alpha \beta})^{\frac{1}{2}}u_{K}^{l}(r)
\hspace*{0.3cm} \cdot
\end{equation}
Here $h_{K}^{\alpha \beta}$ is the norm of the Jacobi polynomial
$P_{K}^{\alpha \beta}(x)$ [24].
The potential matrix element is obtained from eq. (26), using eq. (17) and
eq. (42) of Appendix, in the form
\begin{equation}
V_{KK^{\prime}}(r) = \int_{-1}^{+1}P_{K}^{\alpha \beta}(z)V\left(r\sqrt{\frac{1+z}{2}}\right) 
P_{K^{\prime}}^{\alpha \beta}(z) w_{l}(z) dz ,
\end{equation}
where $w_{l}(z)$ = $(1-z)^{\alpha}$$(1+z)^{\beta}$ is the weight function of the
Jacobi polynomials [24].
For Gaussian interaction with $A$ = 3, the integral can be obtained
analytically [25], from where
one can directly check the numerical accuracy.
\begin{tabbing}
\Large{\bf{III. Numerical method and results}}
\end{tabbing}
\begin{tabbing}
\large{\bf{ A. Numerical method}}
\end{tabbing}
\hspace*{3cm}
For a chosen number of particles $(A)$ and a chosen interaction potential
$(V(r_{ij}))$, we calculate the potential matrix for a fixed value of 
hyperradius $(r)$ from eqs. (29) and (31) 
using a multi-point Gauss-Jacobi quadrature. 
For the present calculation
we select $l$ = 0 and truncate the PH expansion basis of eq.(23) to a maximum
$K$ value $(=K_{max})$. In order to simplify the solution of the set of
coupled differential 
equations, eq. (28), we adopt the hyperspherical adiabatic approximation
(HAA) [16,26]. In this approximation it is assumed that the hyperradial motion is
slow compared to the hyperangular motions. Hence the latter can be solved
adiabatically for a fixed value of $r$ to get an effective potential as
a parametric function of $r$ [16]. This is done by diagonalizing the potential
matrix together with the diagonal hypercentrifugal repulsion and the
trapping potential for each value of $r$ :
\begin{equation}
\sum_{K^{\prime}=1}^{K_{max}} M_{KK^{\prime}}(r) \hspace*{.2cm}
\chi_{K^{\prime}\lambda}(r) = \omega_{\lambda}(r)\chi_{K\lambda}(r)
\end{equation}
where
\begin{equation}
M_{KK^{\prime}}(r) = \overline{V}_{KK^{\prime}}(r) 
+ \left[ \frac{\hbar^{2}}{mr^{2}}\{ {\cal L}({\cal L} +1)
+4K(K+\alpha+\beta+1)\}+V_{trap}(r) \right] \delta_{KK^{\prime}}
\end{equation}
The lowest eigenvalue gives the "lowest eigen potential ", $\omega_{0}(r)$.
As we discussed in the introduction, the hyperradius behaves as the most important collective
coordinate and $\omega_{0}(r)$ is the potential in which the condensate moves as
a "single quantum stuff", except for attractive two-body interactions and $A > A_{cr}$ (see later). Another collective coordinate is the hyperangle $\phi$
appearing in the wavefunction through eqs. (23) and (17), which describes
the deviations of the condensate from hyperspherically symmetric distribution.\\
\hspace*{3cm}
In the HAA approach, an approximate solution of eq. (28) is obtained by
solving a single uncoupled differential equation [16]
\begin{equation}
\left[ -\frac{\hbar^{2}}{m}\frac{d^{2}}{dr^{2}} + \omega_{0}(r)
+ \sum_{K=0}^{K_{max}} |\frac{d\chi_{K0}(r)}{dr}|^{2}-E \right] \zeta_{0}(r)=0
\hspace*{0.3cm} \cdot
\end{equation}
The solution of eq. (34) subject to appropriate boundary conditions
on $\zeta_{0}(r)$ gives the energy $E$, which is an upper bound 
for the eigen energy of eq. (28). The partial waves of eq.(28) are given in
HAA by [16]
\begin{equation}
U_{Kl}(r) \simeq \zeta_{0}(r) \chi_{K0}(r)
\end{equation}
This approximation is usually called uncoupled
adiabatic approximation (UAA) in the literature [16,26]; disregarding the third
term on the left side of eq.(34) one gets the so called extreme adiabatic
approximation (EAA). It has been shown that the HAA is in very good agreement 
(having less than 1\% error) with the exact solution of the CDE for
both atomic [27-29] and nuclear [30-31] cases. Since this is adequate 
for this preliminary application
of this new method, we adopt the HAA, instead of solving
the full set of CDE by exact numerical algorithm like the renormalized Numerov
method [32].
\begin{tabbing}
\large{\bf{ B. Choice of two body interaction potential}}
\end{tabbing}
\hspace*{3cm}
In this report we compare our results with those of the GP equation  
as also 
with other calculations using a contact $\delta$-interaction.
But a $\delta$-function interaction is not a physical one since it
diverges at $r_{ij}$ = 0 and nothing ({\it{e.g.}} centrifugal repulsion)
can prevent its overwhelming effect.
As a result, the Hamiltonian becomes unbound from below for an attractive
$\delta$ interaction. This is manifest in the effective potential $\omega_{0}(r)$,
which for a particle number $(A)$ less than a critical value $(A_{cr})$
produces a local minimum at a finite value of $r$ ( giving rise to a metastable
solution), but $\omega_{0}(r)$ $\rightarrow$  $-\infty$ as $r$ $\rightarrow 0$ 
for any number of particles (see following subsection, as also ref. [12]).
Thus there are no rigorously acceptable and stable solution for any $A$, 
since the attractive essential singularity at $r = 0$ will pull the system to 
$r \rightarrow 0$ and the corresponding wave function will diverge at 
$r = 0$.
Although the $\delta$-function is particularly convenient for analytic 
calculations, it is
desirable to choose an interaction which would either remain finite or at 
worst introduce a removable singularity as $r_{ij}$ $\rightarrow$
$0$ for attractive cases. Then the hyper centrifugal repulsion in eq.(28) (which is non 
vanishing even for $l$=0, $K$ = 0 and increases rapidly as $A$ increases) 
will not allow the 
interacting particles to come too close to each
other. We thus choose a Gaussian potential of strength
$V_{0}$ and range $r_{0}$ 
\begin{equation}
V(r_{ij}) = V_{0} e^{-\frac{r_{ij}^{2}}{r_{0}^{2}}}\hspace*{0.3cm}\cdot
\end{equation}
Choosing appropriate values of $V_{0}$ and $r_{0}$, the potential can be made 
either soft or stiff. A particular experimental situation at the low
temperature limit is characterized by the $s$-wave scattering length
($a_{sc}$). For given values of $V_{0}$ and $r_{0}$, one can calculate
$a_{sc}$ by solving the two-body radial Schr\"odinger equation for positive
energies, in the zero energy limit. Alternately, for a suitably chosen
value of $r_{0}$ and an experimentally known value of $a_{sc}$ one can
find $V_{0}$ numerically from the solution of the two-body Schr\"odinger
equation in the $E \rightarrow 0+$ limit. In Fig. 1, we present a plot of
calculated $a_{sc}$ as a function of $V_{0}$ for $r_{0}$ = 0.0855 $o.u.$.
As is well known, $a_{sc}$ is positive and monotonically
continuous for $V_{0} > 0$. The scattering length becomes negative as $V_{0}$
becomes negative and continues to $-\infty$ at a particular negative 
value of $V_{0}$. At this point, $a_{sc}$ has an infinite discontinuity
and as $V_{0}$ decreases further, $a_{sc}$ starts from + $\infty$ and
decreases continuously to  $- \infty$ at a second particular value of
$V_{0}$. The first, second, ..., branch of the curve (as $V_{0}$ decreases
from positive values) correspond respectively to zero, one, ..., two-body
bound states. For a stable BEC, we choose the first branch of the curve.
From Fig. 1, one notices that for $r_{0}$ = 0.0855 $o.u.$, the first
discontinuity occurs at about $V_{0} = -$ 184 $o.u.$. For $r_{0}$ = 0.005
$o.u.$, this value is much more negative ($- 8.18963 \times 10^{5}$ $o.u.$). In the same figure, we
also plot the Born approximation for $a_{sc}$ (corresponding to $r_{0} =
0.0855$  $o.u.$), given by [12]
\begin{equation}
a_{sc}^{(B)}=\frac{m}{2\pi\hbar^{2}}\int{d^{3}rV(\vec{r})}\hspace*{0.3cm}\cdot
\end{equation}
For a Gaussian interaction this integral can be done analytically and gives
\begin{equation}
a_{sc}^{(B)}=\frac{2m}{\hbar^{2}} V_{0}r_{0}^{3}\frac{\sqrt{\pi}}{4} \cdot
\end{equation} 
From Fig. 1, it is seen that the Born approximation is good only for small
values of $|V_{0}|$. In this work, we use the exact result and not the
Born approximation. For repulsive potentials, we choose a conveniently small
value of $r_{0}$ and calculate $V_{0}$ by the exact procedure.\\
\hspace*{3cm}
Choosing a smaller value of $r_{0}$, $V_{0}$ increases
in magnitude and the potential becomes stiffer. 
For very small values of $r_{0}$, 
$V(r_{ij})$ simulates a $\delta$-function. For attractive interactions, we
perform a model calculation with chosen values of $r_{0}$ and $V_{0}$.

\begin{tabbing}
\large{\bf{ C. Results}}
\end{tabbing}
\hspace*{3cm}
With this choice of potential we have solved the CDE eq.(28) for various
number of particles. We use oscillator units ($o.u.$)
in which energy and length are expressed in units of oscillator energy
and oscillator length ($\hbar\omega$ and $\sqrt{\frac{\hbar}{m\omega}}$ 
respectively, where $\omega$ is the circular frequency of the harmonic
confining potential). The matrix element, eq. (31), has been calculated by
a multi-point Gauss-Jacobi quadrature, the number of points being decided
by the condition of convergence of a typical matrix element. 
We first verify that our results are independent of the choice of $r_{0}$, 
if $V_{0}$ is appropriately calculated using two-body Schr\"odinger
equation, so that $a_{sc}$ has the same value ( 100 
Bohr for $^{87}Rb$, which has a repulsive interaction). 
In a few representative calculations, the ground state energy and low
lying excitation spectrum of the condensate containing $A$ particles have
been found to be stable within numerical errors, for several values of
$r_{0}$ ranging from 0.1 $o.u.$ to 0.005 $o.u$. As for example, the
ground state energy per particle for a condensate containing $A = 10$
bosons approaches a convergence as $r_{0}$ decreases from 0.1 to 0.005. Relative
change in the energy per particle from $r_{0}$ = 0.01 $o.u.$ to 0.005 $o.u.$
is only about 0.012\%.
As $r_{0}$ decreases, the calculation of the  
matrix elements as also the solution of eq. (34) become extremely CPU time 
consuming. This is because for very small $r_{0}$, one has to introduce
very fine $r$-mesh intervals ( typically $10^{-5}$ $o.u.$), which increases
CPU time  
enormously. To keep the numerical calculations manageable, we choose 
$r_{0}$ = 0.005 $o.u.$ and $V_{0} = 3.1985 \times 10^{6}$ $o.u.$  
(which corresponds to JILA 
$^{87}Rb$ experiments with $a_{sc}$ = 100 $Bohr$ and trap frequency $\nu$ =
200 $Hz$). We next test the convergence of our results as $K_{max}$
increases by calculating the ground state energy per particle of the condensate
for $V_{0} = 3.1985 \times 10^{6}$ $o.u.$ and $r_{0}$ = 0.005 $o.u$.
Our results are presented in Table 1. 
It is seen that the energy per particle converges quite rapidly as $K_{max}$
increases. For example, for $A=20$, the change in
energy is less than 0.001\% as $K_{max}$ increases from 2 to 10. 
Another interesting observation is that the ground state energy
{\it{decreases}} as 
$K_{max}$ increases, which is consistent with the Rayleigh-Ritz principle.
Thus it is reassuring that our method is working satisfactorily
and is fast converging.\\ 
\hspace*{3cm}
However a numerical difficulty appears as the particle number ($A$) and
$K_{max}$ 
increase. The quantity $\alpha$ increases rapidly with $A$, (e.g.,
$\alpha$ = 0.5 for $A$ = 3 and 
$\alpha$ = 71 for $A$ = 50), while $\beta$ remains constant at $\frac{1}{2}$ 
(for $l$
=0). Thus the Jacobi polynomial ($P_{n}^{\alpha,\beta}(z)$) as also its weight 
function ($w_{l}(z)$) are
highly asymmetric functions in the interval [-1,1] (see ref. [24]). They
have tremendous  
variation in their
values (e.g. $2^{\alpha}$ to zero) as the argument varies from -1 to +1 
for large $A$. Furthermore $w_{l}(z)$ increases from 0 to $2^{\alpha}$
within a {\it{very small interval}} close to $z=-1$, for large $\alpha$. 
In addition,
$P_{n}^{\alpha,\beta}(z)$ has $n$ nodes in the interval [-1,1].  
Hence unavoidable numerical error creeps into the numerical
integration of the potential matrix, using eq. (31).
Consequently the calculated energy per particle and other physical
quantities show irregularity for $A$ $\geq$ $40$, as also for smaller $A$
with large $K_{max}$. Therefore we have
restricted $A$ to 35. Even for $15 \leq A \leq 35$, some results for large
$K_{max}$ are not reliable. Hence these have been left out in Table 1. In
all subsequent calculations, we keep $K_{max}$ = 4. We are at present 
trying to overcome these difficulties for large $A$ by improved numerical
techniques.\\
\hspace*{3cm}
In Fig. 2, we present a plot of the lowest eigen potential in EAA 
for $A$ = 20, for a model replusive interaction with $V_{0}$ = 20 $o.u.$ and
$r_{0}$ = 0.1 $o.u.$ (dotted curve) corresponding to $a_{sc}$ = 0.01553 $o.u.$
(224.3 $Bohr$). In the 
same figure, we also include the non-interacting  ($V_{0}$ = 0, $a_{sc}$ = 0)
case (continuous curve), which naturally lies below the repulsive
interaction ($a_{sc} > 0$) curve. In Fig. 3, we plot $\omega_{0}(r)$ for an
attractive interaction, {\it{viz.}}, $V_{0} = -$100 $o.u.$, $r_{0}$ =
0.0855 $o.u$ (note from Fig. 1 that this corresponds to zero two-body
bound state and $a_{sc} = - 0.1176$ $o.u.$)
for $A$ = 10. Since we cannot go to large
values of $A$ due to numerical problems mentioned above, we keep 
$A$ small and increase $V_{0}$ to study the
critical behaviour (see below) at a lower value of $A$. Both these curves
have the general features same as those found in earlier calculations using
K-harmonics approximation [13]. Fig. 3 shows a metastable region with a local
minimum of $\omega_{0}(r)$, which is preceded by a collapse region for smaller
$r$. As $A$ increases above a critical value ($A_{cr}$), the metastable region
disappears. This is seen in Fig. 4 for $A$ = 16 for the same $V_{0}$
and $r_{0}$. These features are the same as reported earlier [13]. However,
in our case, since $V(r)$ is finite for $r \rightarrow$ 0, and the repulsive
centrifugal term goes as $\frac{1}{r^{2}}$, there is no real collapse. For
very small $r$, $\omega_{0}(r)$ becomes strongly repulsive even for an
attractive two body interaction. This is represented by the dotted lines in
Figs. 3 and 4. Note that the dotted and continuous parts together constitute the entire calculated $\omega_{0}(r)$ curve. The small $r$ (repulsive) part is plotted with a different (dotted) curve to emphasize that the remaining part (continuous portion) of $\omega_{0}(r)$ has the same behaviour as obtained with attractive contact interaction in ref. [13]. Only the dotted part differs remarkably from the corresponding part in ref. [13]. In reality for $A > A_{cr}$, there is a very narrow and deep well
at a small value of $r$; hence all the particles will be trapped within
this well. As the particles come within a small region, corresponding to a
small value of $r$, the density of the condensate increases, and due to
increased three and higher body collisions, molecule formation takes place
with the disappearance of the BEC. The deep and narrow well in $\omega_{0}(r)$ near the origin, for an attractive two-body interaction with $A > A_{cr}$, can support a lowlying, highly localized bound state, which describes the formation of molecules. Although this is the lowest lying state in the corresponding $\omega_{0}(r)$, it does not represent the ground state of the condensate, which has already "collapsed". This gives a realistic scenario of what
happens as $A$ increases above $A_{cr}$ for attractive interactions. For an
attractive $\delta$-function interaction, the lack of a rigorous solution 
fails to
give a realistic picture and one talks of a "collapse of the condensate" in
a qualitative fashion.\\
\hspace*{3cm}
We next calculate first three excited states for different
number of particles ($A$) in the condensate. These are shown in Fig. 5.
Values of $E_{n}^{ex}$ for $n = 1, 2, 3$ have been represented by
diamonds, pluses and squares respectively. The excitation energy increases
slowly with $A$.
They agree fairly well with the K-harmonic approximation [13]. In Table 2,
we present numerical values and notice that the excitation energies increase
gradually with $A$.\\
\hspace*{3cm}
In Fig. 6, we plot the ground state wave function, $\zeta_{0}(r)$, as a
function of the global length $r$ for various values of $A$. It
is seen that as the particle number increases, the peak of $\zeta_{0}(r)$
shifts towards larger values of $r$. This is understandable, since for
large $A$, the total repulsion of all the pairs increases as $A^{2}$ and
particles are pushed outwards, by the $A$-dependent hypercentrifugal
repulsion in eq. (28).\\
\hspace*{3cm}
Finally we calculate and plot the ground state energy
per particle ($E_{0}/A - \frac{3}{2}\hbar\omega$) as a function of $Aa_{sc}$
for selected values of $A$ (10, 20 and 30) for a repulsive interaction
in Fig. 7. Corresponding curves are from the bottom upwards respectively. We compare these
with the corresponding values calculated from the GP equation. This curve
is the top most in Fig. 7. One notices that our results approach the GP
result as $A$ increases for a fixed $Aa_{sc}$, as expected. We also note
that our energies
are below those of the GP equation, indicating once again a better result from the
variational point of view. 
Fig. 7 agrees
qualitatively with a similar figure of ref. [19], where exact
diagonalization of the Hamiltonian was performed for one and two
dimensional condensates respectively.\\

\begin{tabbing}
{\Large{\bf IV. Conclusions}}
\end{tabbing}
\hspace*{3cm}
In this communication, we have investigated the $T=0$ properties of a
Bose-Einstein condensate (BEC), consisting of $A$ atoms (bosons) trapped
by an external field and interacting {\it{via}} realistic two-body
interactions. An {\it{ab initio}} treatment of the Schr\"odinger equation
involves $3(A-1)$ degrees of freedom for the relative motion. Use of
traditional hyperspherical harmonics expansion (HHE) method is impossible
for $A >$ 3, due to tremendous and mounting complexity of the method as
particle number increases beyond three. We circumvent this difficulty by
exploiting the subset of potential harmonics (PM) basis, instead of the
full set of hyperspherical harmonics (HH) basis. The PH basis is obtained as the
subset of HH needed for expanding the two-body potential for the
interacting pair. The choice of PH
basis corresponds to inclusion of two-body correlations and disregard of
all higher-body correlations in the condensate. On the other hand,
two-body correlations are very important in BEC and cannot be disregarded
as in mean field theories or the GP equation. This assumption is exactly
appropriate for 
the BEC, since for practical realization of BEC, the density of atoms must
be kept so low that there are practically no three and higher body
collisions. Existence of the latter type of collisions would facilitate
formation of molecules and consequent depletion of the condensate. As a
consequence of this assumption, {\it{only four}} active degrees of freedom
of the condensate (instead of a total of $3A-3$ degrees of
freedom for the relative motion of the $A$ particle system) are physically
important - these are constituted by the global length (hyperradius, $r$)
and the three active angle variables of the PH. In effect one
{\it{freezes}} the remaining ($3A-7$) angle variables of PH. This leads to
a tremendous simplification of the actual numerical calculation. Since we
make Faddeev like decomposition of the full wave function, an appropriate
symmetrization of the wave function under exchange of the interacting pair
guarantees full symmetrization. Moreover, the potential matrix elements
involve integrals over {\it{only three}} angle variables, leading to an
immense reduction in the complexity of the numerical procedure for $A$.
Since there are no theoretical restrictions on $A$, this opens the
possibility of an approximate but very reliable, {\it{ab initio}} solution
of the large but finite body condensate. However, a numerical difficulty
arises due to the fact that the parameter $\alpha$  $(=(3A-8)/2)$ of the
Jacobi polynomials, $P^{\alpha,\beta}_{n}(x)$, and its associated weight
function, become very large as $A$ increases. These cause numerical problems,
for $A \geq 40$. We are at present attempting to remove this difficulty by
appropriate numerical procedure. In the present report, we restrict
ourselves to $A \leq 35$, for which reliable calculations are possible.\\
\hspace*{3cm}
We have compared our results with earlier calculations for $A=3$ [12],
K-harmonic approximation [13], exact diagonalization of the Hamiltonian in
one and two dimensions [19] as also with the predictions of the GP
equation [8]. As a preliminary calculation we have taken two-body Gaussian
interactions of varying range. Our results agree qualitatively with the
previous ones, most of which use a contact interaction. This demonstrates
the reliability and feasibility of our method. Thus a reliable {\it{ab
initio}} calculation for a large but finite number of atoms in a
condensate, where individual particles interact via realistic two-body
interactions, appears feasible. Extension of our method to larger number
of particles as also use of more realistic two-body interaction is
underway.\\[1cm]
\begin{tabbing}
\large{\bf{Acknowledgements}}\\
\end{tabbing}
This work has been supported by a grant from the Department of Science and
Technology (DST), Government of India under a research project. One of the
authors (BC) wishes to thank Prof. D. K. Watson for providing a
Post-doctoral fellowship at the University of Oklahoma (U.S.A.), where
part of the work was done. She also wishes to thank Dr. B. A. McKinney for
providing the code for solving the GP equation.\\
\newpage
\begin{center}
{\Large{\bf{Appendix}}}\\[0.5cm]
{\Large{\bf{Hyperspherical variables and hyperspherical harmonics}}}\\
\end{center}
{\large{\bf{ A1 : Hyperspherical variables}}}\\[0.5cm]
\hspace*{3cm} The relative motion of the $A=(N+1)$ particle system is described
in terms of $N$ Jacobi coordinates defined by eq. (8) and having $3N$
degrees of freedom. An equivalent set of hyperspherical variables is constituted
by the hyperradius $(r)$ defined by eq.(11), $2N$ spherical polar angles of 
$\vec{\zeta_{1}},\vec{\zeta_{2}}.....,\vec{\zeta_{N}}$ and
$(N-1)$ hyperangles
$\{\phi_{2},\phi_{3},....\phi_{N}\}$ giving the length of the Jacobi vectors
$\vec{\zeta_{1}},\vec{\zeta_{2}}.....,\vec{\zeta_{N}}$, through
\begin{equation}
\begin{array}{rcl}
\zeta_{N} &=& r \hspace*{.5cm}cos \phi_{N}\\
\zeta_{N-1} &=& r  \hspace*{.5cm} sin \phi_{N}  \hspace*{.5cm}cos \phi_{N-1}\\
\zeta_{N-2} & =& r  \hspace*{.5cm} sin \phi_{N} \hspace*{.5cm}sin
\phi_{N-1}  \hspace*{.5cm} cos \phi_{N-2}\\
& .&\\
&.&\\
&.&\\
\zeta_{2} &=& r  \hspace*{.5cm}sin \phi_{N}  \hspace*{.5cm} sin \phi_{N-1}....
sin\phi_{3} \hspace*{.5cm} cos \phi_{2}\\
\zeta_{1} &=& r  \hspace*{.5cm} sin \phi_{N} \hspace*{.5cm}sin \phi_{N-1}
....sin \phi_{3} \hspace*{.5cm}sin \phi_{2}\\
\hspace*{5cm} \left(\phi_{1}=0\right)
\end{array}
\end{equation}
Eq. (39) automatically satisfies eq. (11).\\[2cm]
\begin{tabbing}
{\large{\bf{A2. Grand orbital operator}}}
\end{tabbing}
\hspace*{3cm}
The general grand orbital operator, $L^{2}(\Omega_{N})$ of eq. (15) is defined
through [14]
\begin{equation}
\begin{array}{rcl}
L_{i}^{2}(\Omega_{i})&=&\frac{\partial^{2}}{\partial
\phi_{i}^{2}}+\left[3(i-2)cot \phi_{i}+2(cot
\phi_{i}-tan \phi_{i})\right]\frac{\partial}{\partial \phi_{i}}
+\frac{l^{2}(\omega_{i})}{cos^{2}\phi_{i}}
+\frac{L^{2}_{i-1}(\Omega_{i-1})}{sin^{2}\phi_{i}}\\
&=& 4(1-z_{i}^{2})\frac{\partial^{2}}{\partial z_{i}^{2}}
+6[2-i(1+z_{i})]\frac{\partial}{\partial z_{i}}
+2 \frac{l^{2}(\omega_{i})}{1+z_{i}}
+2 \frac{L^{2}_{i-1}(\Omega_{i-1})}{1-z_{i}}\\
&&  (i= 2,..........N)
\end{array}
\end{equation}
 where $z_{i}$ = $cos 2\phi_{i}$, $\omega_{i}$ represents the set of two
 polar angles of $\vec{\zeta_{i}}$ and $\phi_{i}$'s are given by eq.
(39). Note that $L_{1}^{2}(\Omega_{1})$ = $l^{2}(\omega_{1})$ and 
$L_{N}^{2}(\Omega_{N})$  $\equiv$ $L^{2}(\Omega_{N})$ appear in eq. 
(15). \\[2cm]
\begin{tabbing}
{\large{\bf{ A3. Hyperspherical harmonics}}}
\end{tabbing}
\hspace*{3cm}
An eigenfunction of $L^{2}(\Omega_{N})$ is called hyperspherical 
harmonics (HH) and is given (without angular momentum coupling) by [34]
\begin{equation}
Y_{[{\cal L}]}(\Omega_{N})=Y_{l_{1}m_{1}}(\omega_{1})
\prod_{j=2}^{N}Y_{l_{j}m_{j}}(\omega_{j})
^{(j)}P_{{\cal L}_{j}}^{l_{j},{\cal L}_{j-1}}(\phi_{j})
\end{equation}
where
\begin{equation}
\begin{array}{rcl}
^{(j)}P_{{\cal L}_{j}}^{l_{j},{\cal L}_{j-1}}(\phi_{j}) &=& \{\frac{2
\nu_{j}\Gamma(\nu_{j}-n_{j})\Gamma(n_{j}+1)}
{\Gamma(\nu_{j}-n_{j}-l_{j}-\frac{1}{2})
\Gamma(n_{j}+l_{j}+\frac{3}{2})}\}^{\frac{1}{2}}\\
&& (cos \phi_{j})^{l_{j}}(sin
\phi_{j})^{{\cal L}_{j-1}}P_{n_{j}}^{\nu_{j-1},l_{j}+\frac{1}{2}}(cos 2\phi_{j})
 \hspace*{0.5cm}(j=2,3,...,N)
\end{array}
\end{equation}
with
\begin{equation}
\begin{array}{rcl}
\nu_{j} &=&  \nu_{j-1} + 2 n_{j} + l_{j} + \frac{3}{2}\\
  &=&   {\cal L}_{j} + \frac{3j}{2} - 1\\
&=&    {\cal L}_{j-1} + 2n_{j} + l_{j} + \frac{3j}{2} -1\\
& &  \hspace*{3cm} (j= 2,3,...,N)\\
\end{array}
\end{equation}
In eq. (42) $P_{n}^{\alpha,\beta}(x)$ is a Jacobi Polynomial. 
In eq. (41), $[{\cal L}]$ representa the set of quantum
numbers $ \{ (l_{1},m_{1}), \hspace*{.2cm} (l_{2},m_{2}), \hspace*{.2cm}
..., (l_{N},m_{N}), n_{2}, n_{3}, ..., n_{N} \}$ for a fixed value of grand orbital 
quantum number ${\cal L}$ = $ {\cal L}_{N}$. 
The quantum number ${\cal L}_{i}$ is defined through 
\begin{equation}
{\cal L}_{i} = {\cal L}_{i-1} + 2 n_{i} + l_{i}
\end{equation}
with ${\cal L}_{1}$ = $l_{1}$. Hence
\begin{equation}
{\cal L} \equiv {\cal L}_{N} = l_{1} + \sum_{j=2}^{N} (2n_{j}+l_{j})
\end{equation}
The HH of eq. (41) forms the uncoupled basis.
For systems with a good orbital angular momemtum 
$\vec{L}$ = $\vec{l_{1}} + \vec{l_{2}}+ ... + \vec{l_{N}}$, one has to couple
the individual orbital angular momenta - then the projection quantum numbers
$m_{1}$, $m_{2}$, ..., $m_{N}$ are replaced by the $(N-1)$ intermediately
coupled angular momenta and the projection $M$ of $\vec {L}$. \\
\hspace*{3cm}
The potential harmonics (PH) given by eq. (17) corresponds to $l_{N}$ = $l$,
$l_{1}$ = $l_{2}$ = $l_{3}$ =  ...  = $l_{N-1}$ = 0, such that $L$ = $l_{N}$ = $l$, 
$M$ = $m_{N}$ = $m$ and grand orbital ${\cal L}$ $\equiv$ ${\cal L}_{N}$ = $2K+l$ with 
$n_{2}$ = $n_{3}$ =  ...  = $n_{N-1}$ = 0 and $K$ = $n_{N}$. Substitution
of these in eqs. (41) - (43) gives the PH of eq. (17).

\newpage
Table 1.
Calculated ground state energy per particle (in $o.u.$) of the condensate
containing $A$ particles for various $K_{max}$ values, showing convergence
trend as $K_{max}$ increases 
($V_{0} = 3.1985 \times 10^{6}$ $o.u.$ and $r_{0} = 0.005$ $o.u.$)
\begin{center}
\begin{tabular}{|l|l|l|l|l|l|l|}
\hline
$K_{max}$ & 2 & 4 & 6 & 8 & 10 & 12\\
$A$ & & & & & &\\
\hline
3 & 1.50041 & 1.50031 & 1.50026 & 1.50023 & 1.50021 & 1.50019\\
5 & 1.50123 & 1.50117 & 1.50112 & 1.50108 & 1.50104 & 1.50101\\
10 & 1.50350 & 1.50348 & 1.50346 & 1.50344 & 1.50342 & 1.50340\\
15 & 1.50453 & 1.50451 & 1.50450 & 1.50449 & 1.50449 & \\
20 & 1.50539 & 1.50538 & 1.50537 & 1.50536 & 1.50536 & \\
25 & 1.50618 & 1.50617 & 1.50617 & 1.50616 & 1.50616 & \\
30 & 1.50693 & 1.50692 & 1.50692 & 1.50691 & & \\
35 & 1.50764 & 1.50763 & 1.50763 & 1.50766 & & \\
\hline
\end{tabular}
\end{center}
\vskip5pt
Table 2.
Calculated excitation energies (in $o.u.$) of the first three excited states
for different numbers ($A$) of
$^{87}Rb$ atoms (parameters as in Table. 1).\\[0.5cm]

\begin{center}
%
\\[1cm]
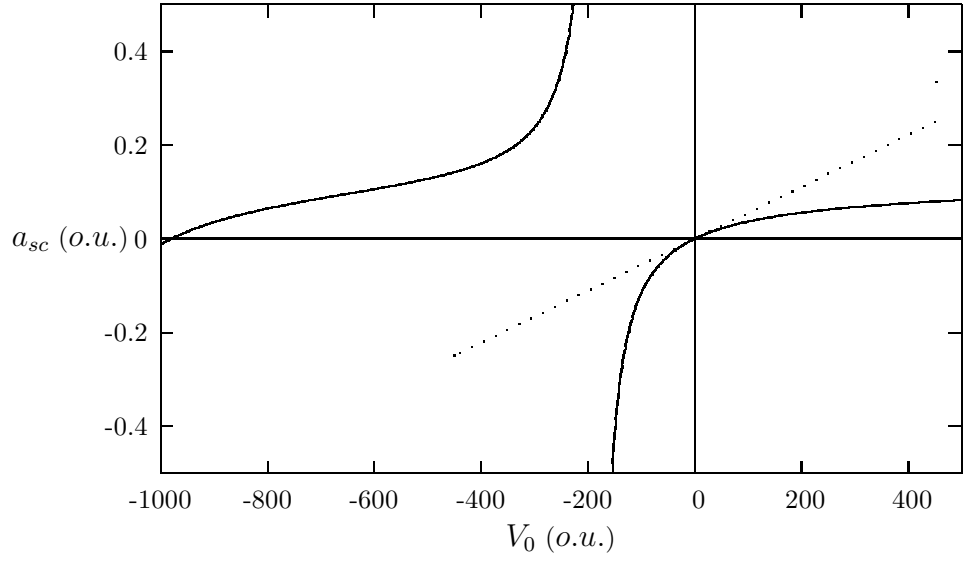
Fig. 1 - Plot of calculated $a_{sc}$ as a function of $V_{0}$ for $r_{0} =
0.0855$ $o.u.$. The dotted line corresponds to the Born approximation
($a_{sc}^{(B)}$).\\

\setlength{\unitlength}{0.240900pt}
\ifx\plotpoint\undefined\newsavebox{\plotpoint}\fi
\sbox{\plotpoint}{\rule[-0.200pt]{0.400pt}{0.400pt}}%

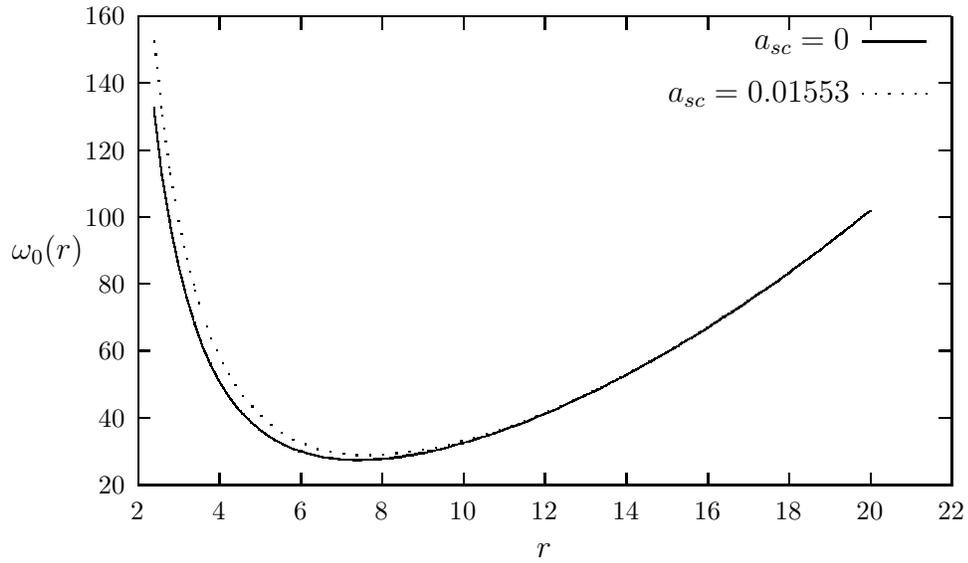
Fig. 2 - Lowest eigen potential for $A=20$ as a function of $r$. 
Continuous curve is for $a_{sc}=0$ (no two-body interaction) and the 
dotted curve is for a repulsive interaction ($a_{sc}=0.01553$ $o.u.$).\\
 
\setlength{\unitlength}{0.240900pt}
\ifx\plotpoint\undefined\newsavebox{\plotpoint}\fi
%
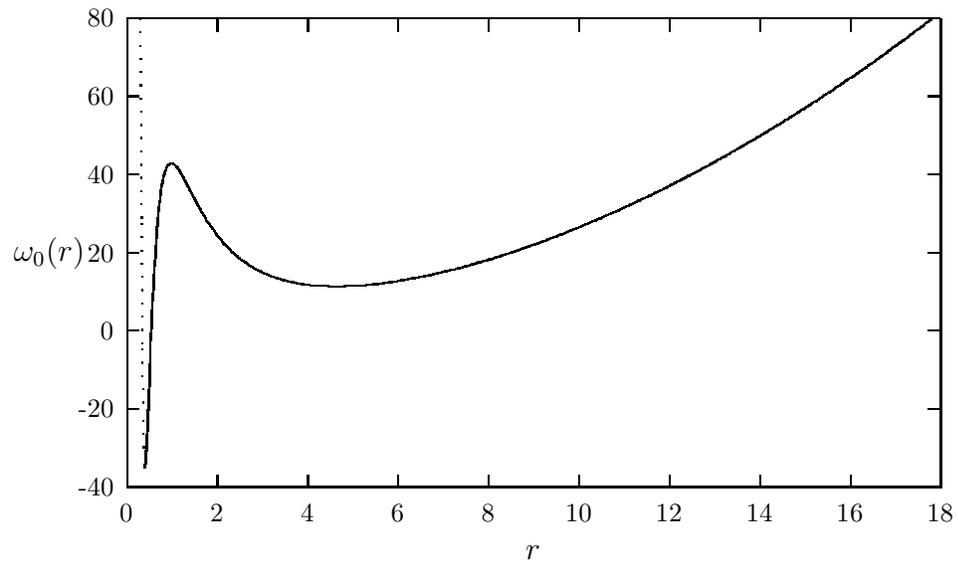
Fig. 3 - Plot of $\omega_{0}(r)$ as a function of $r$ (dotted and
continuous curves together) for $A=10$ (subcritical
number) for a model attractive two-body interaction ($V_{0} = -100$ $o.u.$, 
$r_{0}$ = 0.0855 $o.u.$), which corresponds to $a_{sc} = -0.1176$ $o.u.$\\

\setlength{\unitlength}{0.240900pt}
\ifx\plotpoint\undefined\newsavebox{\plotpoint}\fi

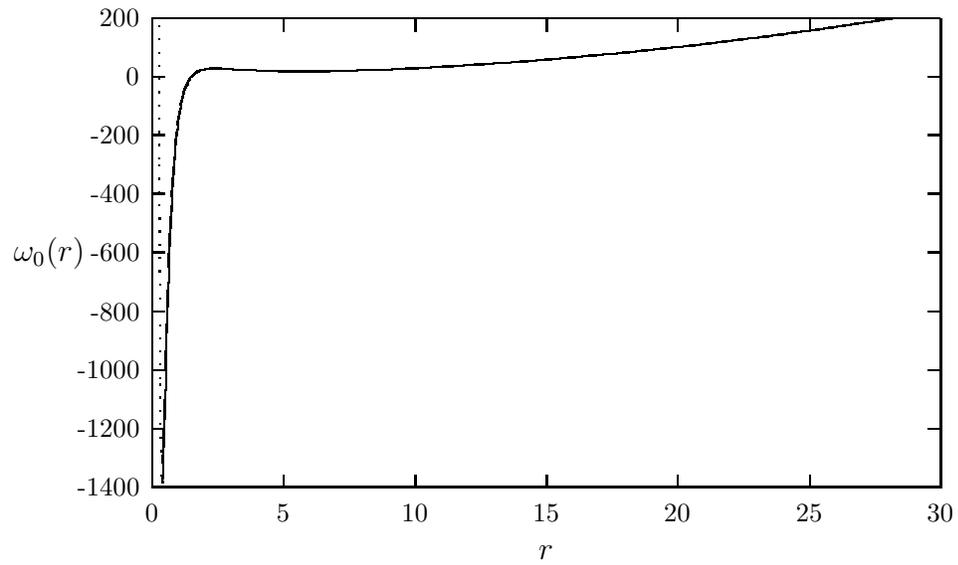
Fig. 4 - Plot of $\omega_{0}(r)$ as a function of $r$ (dotted and
continuous curves together) for $A = 16$ 
(critical number) for the same attractive two-body potential as in Fig. 3.\\  
\setlength{\unitlength}{0.240900pt}
\ifx\plotpoint\undefined\newsavebox{\plotpoint}\fi
\sbox{\plotpoint}{\rule[-0.200pt]{0.400pt}{0.400pt}}%
\begin{picture}(1500,900)(0,0)
\font\gnuplot=cmr10 at 10pt
\gnuplot
\sbox{\plotpoint}{\rule[-0.200pt]{0.400pt}{0.400pt}}%
\put(141.0,197.0){\rule[-0.200pt]{4.818pt}{0.400pt}}
\put(121,197){\makebox(0,0)[r]{ 2}}
\put(1419.0,197.0){\rule[-0.200pt]{4.818pt}{0.400pt}}
\put(141.0,344.0){\rule[-0.200pt]{4.818pt}{0.400pt}}
\put(121,344){\makebox(0,0)[r]{ 3}}
\put(1419.0,344.0){\rule[-0.200pt]{4.818pt}{0.400pt}}
\put(141.0,492.0){\rule[-0.200pt]{4.818pt}{0.400pt}}
\put(121,492){\makebox(0,0)[r]{ 4}}
\put(1419.0,492.0){\rule[-0.200pt]{4.818pt}{0.400pt}}
\put(141.0,639.0){\rule[-0.200pt]{4.818pt}{0.400pt}}
\put(121,639){\makebox(0,0)[r]{ 5}}
\put(1419.0,639.0){\rule[-0.200pt]{4.818pt}{0.400pt}}
\put(141.0,786.0){\rule[-0.200pt]{4.818pt}{0.400pt}}
\put(121,786){\makebox(0,0)[r]{ 6}}
\put(1419.0,786.0){\rule[-0.200pt]{4.818pt}{0.400pt}}
\put(222.0,123.0){\rule[-0.200pt]{0.400pt}{4.818pt}}
\put(222,82){\makebox(0,0){ 5}}
\put(222.0,840.0){\rule[-0.200pt]{0.400pt}{4.818pt}}
\put(425.0,123.0){\rule[-0.200pt]{0.400pt}{4.818pt}}
\put(425,82){\makebox(0,0){ 10}}
\put(425.0,840.0){\rule[-0.200pt]{0.400pt}{4.818pt}}
\put(628.0,123.0){\rule[-0.200pt]{0.400pt}{4.818pt}}
\put(628,82){\makebox(0,0){ 15}}
\put(628.0,840.0){\rule[-0.200pt]{0.400pt}{4.818pt}}
\put(831.0,123.0){\rule[-0.200pt]{0.400pt}{4.818pt}}
\put(831,82){\makebox(0,0){ 20}}
\put(831.0,840.0){\rule[-0.200pt]{0.400pt}{4.818pt}}
\put(1033.0,123.0){\rule[-0.200pt]{0.400pt}{4.818pt}}
\put(1033,82){\makebox(0,0){ 25}}
\put(1033.0,840.0){\rule[-0.200pt]{0.400pt}{4.818pt}}
\put(1236.0,123.0){\rule[-0.200pt]{0.400pt}{4.818pt}}
\put(1236,82){\makebox(0,0){ 30}}
\put(1236.0,840.0){\rule[-0.200pt]{0.400pt}{4.818pt}}
\put(1439.0,123.0){\rule[-0.200pt]{0.400pt}{4.818pt}}
\put(1439,82){\makebox(0,0){ 35}}
\put(1439.0,840.0){\rule[-0.200pt]{0.400pt}{4.818pt}}
\put(141.0,123.0){\rule[-0.200pt]{312.688pt}{0.400pt}}
\put(1439.0,123.0){\rule[-0.200pt]{0.400pt}{177.543pt}}
\put(141.0,860.0){\rule[-0.200pt]{312.688pt}{0.400pt}}
\put(40,491){\makebox(0,0){$E_{n}^{ex}$}}
\put(790,21){\makebox(0,0){$A$}}
\put(141.0,123.0){\rule[-0.200pt]{0.400pt}{177.543pt}}
\put(141,197){\raisebox{-.8pt}{\makebox(0,0){$\Diamond$}}}
\put(222,197){\raisebox{-.8pt}{\makebox(0,0){$\Diamond$}}}
\put(425,197){\raisebox{-.8pt}{\makebox(0,0){$\Diamond$}}}
\put(628,197){\raisebox{-.8pt}{\makebox(0,0){$\Diamond$}}}
\put(831,197){\raisebox{-.8pt}{\makebox(0,0){$\Diamond$}}}
\put(1033,198){\raisebox{-.8pt}{\makebox(0,0){$\Diamond$}}}
\put(1236,202){\raisebox{-.8pt}{\makebox(0,0){$\Diamond$}}}
\put(1439,209){\raisebox{-.8pt}{\makebox(0,0){$\Diamond$}}}
\put(141,492){\makebox(0,0){$+$}}
\put(222,492){\makebox(0,0){$+$}}
\put(425,493){\makebox(0,0){$+$}}
\put(628,493){\makebox(0,0){$+$}}
\put(831,494){\makebox(0,0){$+$}}
\put(1033,496){\makebox(0,0){$+$}}
\put(1236,505){\makebox(0,0){$+$}}
\put(1439,523){\makebox(0,0){$+$}}
\sbox{\plotpoint}{\rule[-0.400pt]{0.800pt}{0.800pt}}%
\put(141,787){\raisebox{-.8pt}{\makebox(0,0){$\Box$}}}
\put(222,788){\raisebox{-.8pt}{\makebox(0,0){$\Box$}}}
\put(425,788){\raisebox{-.8pt}{\makebox(0,0){$\Box$}}}
\put(628,791){\raisebox{-.8pt}{\makebox(0,0){$\Box$}}}
\put(831,793){\raisebox{-.8pt}{\makebox(0,0){$\Box$}}}
\put(1033,805){\raisebox{-.8pt}{\makebox(0,0){$\Box$}}}
\put(1236,812){\raisebox{-.8pt}{\makebox(0,0){$\Box$}}}
\put(1439,826){\raisebox{-.8pt}{\makebox(0,0){$\Box$}}}
\end{picture}\\[1cm]
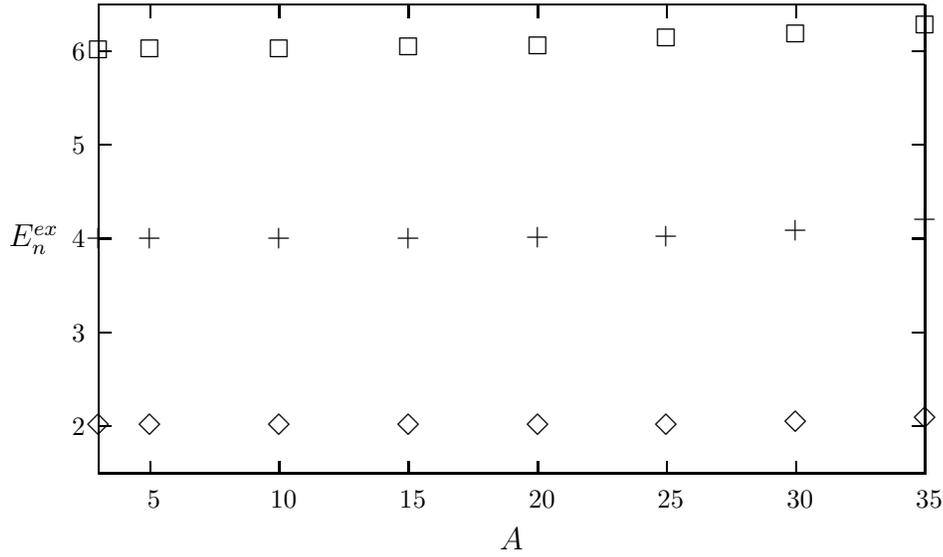
Fig. 5 - Three low-lying excitation frequencies for various values of 
particle number $(A)$, corresponding to the JILA experiment with $^{87}Rb$ 
atoms ( $a_{sc}$ = 100 $Bohr$, oscillator frequency = 200 $Hz$).
Energies are in oscillator units.\\
\setlength{\unitlength}{0.240900pt}
\ifx\plotpoint\undefined\newsavebox{\plotpoint}\fi

\\
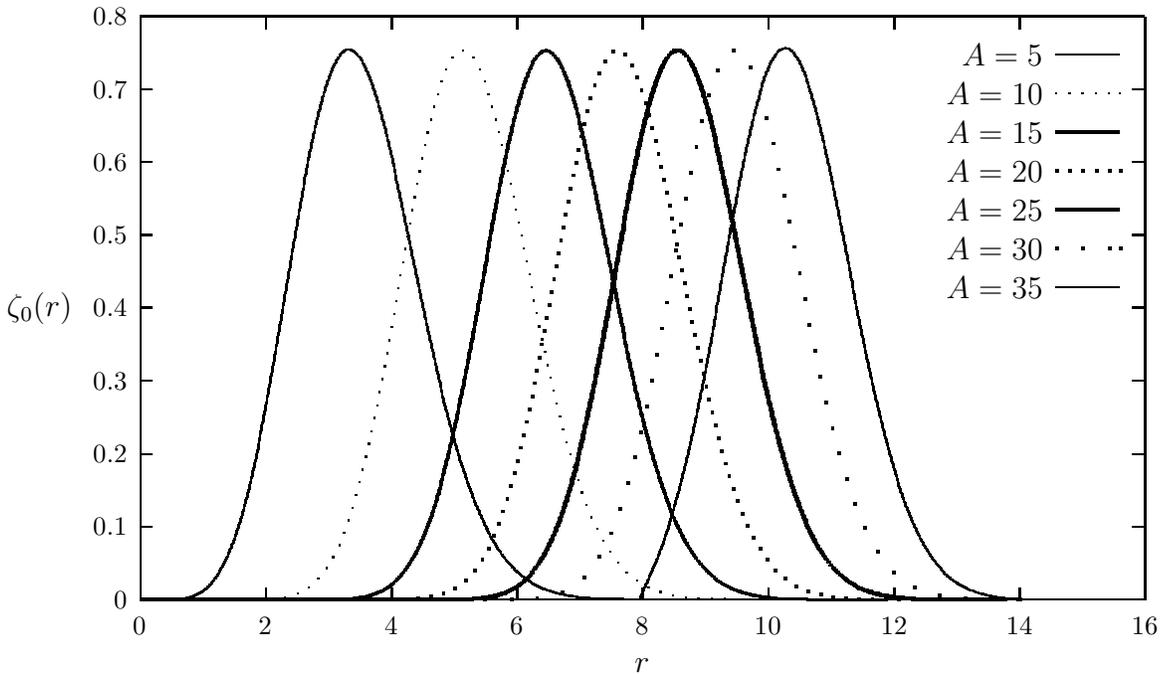
Fig. 6 - Plot of ground state wave function as a function of hyperradius 
($r$) for various indicated value of $A$, in the chosen trap.\\

\setlength{\unitlength}{0.240900pt}
\ifx\plotpoint\undefined\newsavebox{\plotpoint}\fi
\sbox{\plotpoint}{\rule[-0.200pt]{0.400pt}{0.400pt}}%
%
\\[1cm]
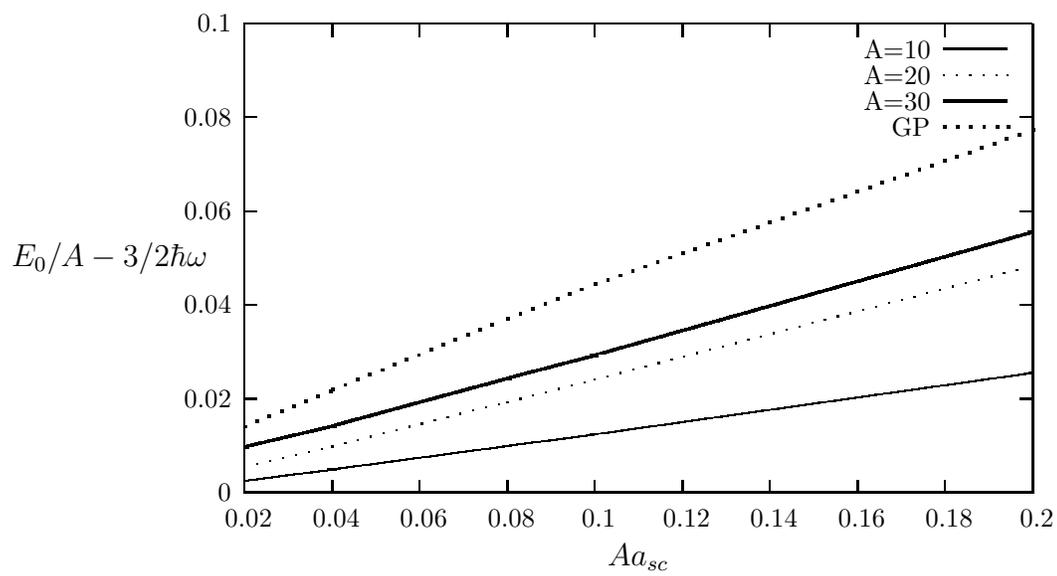
Fig. 7 - Plot of ground state energy per particle $(E_{0}/A-3/2 \hbar\omega)$ 
as a function of $Aa_{sc}$ for a repulsive interaction for indicated 
values of $A$ and the GP results. 
\end{document}